**A distributed regression analysis application based on SAS software**

**Part I: Linear and logistic regression**

**Keywords:** Distributed regression analysis, Distributed data networks, Privacy-protecting methods, PopMedNet[TM]

**Authors**
§ Corresponding author
**Qoua L. Her, PharmD, MSPharm, MSc**
Department of Population Medicine
Harvard Medical School and Harvard Pilgrim Health Care Institute
401 Park Street, Suite 401 East
Boston, MA 02215
Qoua_Her@harvardpilgrim.org
617-867-4885

**Yury Vilk, PhD**
Department of Population Medicine
Harvard Medical School and Harvard Pilgrim Health Care Institute
401 Park Street, Suite 401 East
Boston, MA 02215
Yury_Vilk@harvardpilgrim.org

**Jessica Young, PhD**
Department of Population Medicine
Harvard Medical School and Harvard Pilgrim Health Care Institute
401 Park Street, Suite 401 East
Boston, MA 02215
jyoung@hsph.harvard.edu

**Zilu Zhang, MSc**
Department of Population Medicine
Harvard Medical School and Harvard Pilgrim Health Care Institute
401 Park Street, Suite 401 East
Boston, MA 02215
Zilu_Zhang@harvardpilgrim.org

**Jessica M. Malenfant, MPH**
Department of Population Medicine
Harvard Medical School and Harvard Pilgrim Health Care Institute
401 Park Street, Suite 401 East
Boston, MA 02215
Jessica_Malenfant@harvardpilgrim.org

**Sarah Malek, MPPA**
Department of Population Medicine

Harvard Medical School and Harvard Pilgrim Health Care Institute
401 Park Street, Suite 401 East
Boston, MA 02215
Sarah_Malek@harvardpilgrim.org

**Sengwee Toh, ScD**
Department of Population Medicine
Harvard Medical School and Harvard Pilgrim Health Care Institute
401 Park Street, Suite 401 East
Boston, MA 02215
Darren_Toh@harvardpilgrim.org

**Word Count**: 8,532

# ABSTRACT

Previous work has demonstrated the feasibility and value of conducting distributed regression analysis (DRA), a privacy-protecting analytic method that performs multivariable-adjusted regression analysis with only summary-level information from participating sites. To our knowledge, there are no DRA applications in SAS, the statistical software used by several large national distributed data networks (DDNs), including the Sentinel System and PCORnet. SAS/IML is available to perform the required matrix computations for DRA in the SAS system. However, not all data partners in these large DDNs have access to SAS/IML, which is licensed separately. In this first article of a two-paper series, we describe a DRA application developed for use in Base SAS and SAS/STAT modules for linear and logistic DRA within horizontally partitioned DDNs and its successful tests.

*Word Count (max. 250): 125*

**Funding Statement:** This work was supported by the Office of the Assistant Secretary for Planning and Evaluation (ASPE) and the Food and Drug Administration (HHSF223201400030I/HHSF22301006T).

**Competing Interests Statement:** Dr. Toh is Principal Investigator of projects funded by the National Institutes of Health (U01EB023683) and the Patient-Centered Outcomes Research Institute (ME-1403-11305).

**Contributorship Statement:** All authors contributed to the conception, design, analysis, and interpretation of this study. QLH, YV, JY, and ST led the drafting of the manuscript and revising it for critical important intellectual content. JMM, SM, and ZZ contributed to the conception and design, drafting of the manuscript, or revising it. All authors commented on manuscript drafts and gave their approval for the final version to be published. ST obtained funding and was responsible for supervision of all activities.

# 1. Introduction

Electronic data collected as part of real-world healthcare delivery is used in research and public health surveillance to generate evidence that can improve health at the population- and individual-level. It is common to pool individual-level data from multiple sources to increase sample size and improve generalizability of the study findings. However, sharing detailed individual-level information raises concerns about individual privacy and confidentiality, which may deter multi-center collaborations (Maro et al. 2009; Brown et al. 2010; Toh et al. 2011). Data organized in a distributed data network (DDN), where data remain behind each data partner's firewall, alleviates some of these concerns (Diamond, Mostashari, and Shirky 2009; Maro et al. 2009; Brown et al. 2010; Toh et al. 2011). Several analytic methods are available to perform statistical analysis within DDNs, but methods that only require summary-level information are increasingly preferred because they offer additional privacy protection (Toh et al. 2011; Rassen et al. 2013). Distributed regression analysis (DRA) is one such method that allows multivariable regression analysis using only summary-level information and has been shown to produce results equivalent within machine precision to those from pooled individual-level data analysis (Karr et al. 2004; Fienberg et al. 2006; Wu et al. 2012; Wolfson et al. 2010; Toh et al. 2014; Dankar 2015).

There are currently a number of R-based software applications that allow users to perform DRA (Wolfson et al. 2010; Jiang et al. 2013; Lu et al. 2015; Meeker et al. 2015; Narasimhan et al. 2017). To our knowledge, there are no DRA applications in SAS, the statistical software used by several existing national DDNs in the United States, including the Sentinel System (a DDN funded by the U.S. Food and Drug Administration to conduct medical product safety surveillance) (Platt et al. 2012; Ball et al. 2016) and PCORnet (a DDN funded by the Patient-Centered Outcomes Research Institute to perform comparative effectiveness research) (Fleurence et al. 2014).

This is the first article of a two-paper series. In this article, we describe a DRA application in SAS for linear and logistic DRA within horizontally partitioned DDNs, a setting where different databases contain information about different individuals. We describe Cox proportional hazards DRA in our companion paper. The DRA application comprises two interlinked packages of SAS macros and programs – one for the analysis center and one for the data-contributing sites (i.e., data partners). A key advantage of our DRA application is that it requires only Base SAS and SAS/STAT modules. While SAS/IML is available to perform the required matrix computations, not all data partners have SAS/IML as it is licensed separately. Another advantage is that we have fully integrated the DRA application with PopMedNet$^{TM}$, an open-source query distribution software that supports automatable file transfer between the analysis center and data partners (Her et al. 2018). This ability to automate information exchange during the iterative process of model fitting substantially improves the feasibility of using the DRA application in large-scale multi-center studies.

We organize the article as follows. In Section 2, we describe our distributed implementation of the iteratively reweighted least squares (IRLS) algorithm for generalized linear models (GLMs) using only Base SAS and SAS/STAT to produce parameter estimates, standard errors, and goodness-of-fit measures. In Section 3, we explain how to use our packages at the analysis center and data partner sites to perform DRA. Specifically, we describe the structure and user-specified parameters of the master programs (wrappers) and the main macro %*distributed_regression*, and how to execute these programs, both manually and automatically via PopMedNet. In Section 4, we present the results from the DRA application in several empirical examples and compare them with the results obtained from standard SAS procedures that analyzed pooled individual-level datasets. In Section 5, we discuss possible extensions of the DRA application.

## 2. Distributed regression analysis for horizontally partitioned data

### 2.1 Overview

We consider a horizontally partitioned DDN where each data partner holds distinct individual cohorts. Numerous secure multiparty computation protocols have been presented in the literature for this distributed data environment (Karr et al. 2009; Dankar 2015). Our implementation uses a secure protocol with a semi-trusted third party as the analysis center. We define a semi-trusted third party as any party that data partners trust with summary-level data but not with individual-level data, and that does not share data from one data partner with another without explicit consent from the data partner. The semi-trusted third party can itself be a data-contributing site. Below we describe our computational algorithm to implement DRA for fitting GLMs using only the Base SAS and SAS/STAT modules.

### 2.2 Distributed iterative reweighted least squares

Linear and logistic regression models, along with other commonly used models such as Poisson regression, are special cases of GLMs (McCullagh and Nelder 1989). Maximum likelihood estimators of GLM regression coefficients can be obtained using an IRLS algorithm. When the link function for the GLM is chosen as the canonical link, this is equivalent to the Newton-Raphson algorithm. In this section, we describe a distributed version of the IRLS algorithm for GLMs such that individual-level data from a given site does not need to be shared with other sites nor with the analysis center.

Let $K$ denote the number of sites, and $n_k$ the number of individuals at site $= 1, \ldots, K$. Further, let $(Y_{i,k}, \mathbf{X}_{i,k}, w_{i,k}), i = 1, \ldots, n_k$, denote the observed data for individuals $i$ at site $k$, with $Y_{i,k}$ the outcome, $\mathbf{X}_{i,k}$ a vector of $p$-covariate values for individual $i$, and $w_{i,k}$ an individual-level weight. Let $\mathbf{Z}_{i,k} = \mathbf{1} || \mathbf{X}_{i,k}$ and $N = \sum_{k=1}^{K} n_k$ denote the sum of all observations. The input dataset at site $k$ has the following structure:

$$
\begin{matrix}
w_{1,k} & X_{1,k,1} & \ldots & X_{1,k,p} & Y_{1,k} \\
\vdots & \vdots & \vdots & \vdots & \vdots \\
w_{N,k} & X_{N,k,1} & \ldots & X_{N,k,p} & Y_{N,k}
\end{matrix}
\quad (1)
$$

A GLM assumes that $Y_{i,k}$ is distributed according to an exponential family (e.g., normal, binomial, Poisson) with:

$$E[Y_{i,k}|\mathbf{Z}_{i,k}] = \mu(\boldsymbol{\beta}^T \mathbf{Z}_{i,k})$$

$$var[Y_{i,k}|\mathbf{Z}_{i,k}] = v(\boldsymbol{\beta}^T\mathbf{Z}_{i,k})$$

where $\boldsymbol{\beta}$ is a $p + 1$ length vector of unknown regression coefficients.

### 2.2.1 Special case of linear regression

Before we consider how to estimate $\boldsymbol{\beta}$ generally via IRLS in this setting, we first consider the special case where we select the GLM as a linear regression model, i.e., where $Y_{i,k}$ is assumed to follow a normal distribution with $\mu(\boldsymbol{\beta}^T\mathbf{Z}_{i,k}) = \boldsymbol{\beta}^T\mathbf{Z}_{i,k}$ and $(\boldsymbol{\beta}^T\mathbf{Z}_{i,k}) = v = \sigma^2$. It follows from standard theory that a maximum likelihood estimate of $\boldsymbol{\beta}$ in this special case can be obtained by solving the (possibly weighted) least squares equations:

$$\sum_{k=1}^{K}\sum_{i=1}^{n_k} w_{i,k}(Y_{i,k} - \boldsymbol{\beta}^T\mathbf{Z}_{i,k})\mathbf{Z}_{i,k} = 0 \tag{2}$$

with respect to $\boldsymbol{\beta}$. In this case, an exact solution exists which is:

$$\widehat{\boldsymbol{\beta}} = \left(\sum_{k=1}^{K}\mathbf{Z}_k^T\mathbf{W}_k\mathbf{Z}_k\right)^{-1}\left(\sum_{k=1}^{K}\mathbf{Z}_k^T\mathbf{W}_k\mathbf{Y}_k\right) \tag{3}$$

with $\mathbf{Y}_k$ representing a vector of length $n_k$ with elements $Y_{i,k}$, $\mathbf{Z}_k$ a matrix of dimension $n_k * (p + 1)$ with rows $\mathbf{Z}_{i,k}$ and $\mathbf{W}_k$ a diagonal matrix of dimension $n_k * n_k$ with diagonal elements $w_{i,k}$ $i = 1, \dots n_k$. Importantly, the matrices $\mathbf{Z}_k^T\mathbf{W}_k\mathbf{Z}_k$ and $\mathbf{Z}_k^T\mathbf{W}_k\mathbf{Y}_k$ can be calculated separately at each site $k$. These matrices are highly summarized and can be brought to the analysis center with lower privacy risk because the dimension of $\mathbf{Z}_k^T\mathbf{W}_k\mathbf{Z}_k$ is $(p + 1) * (p + 1)$, which is much smaller than the dimension of individual-level matrix $\mathbf{Z}_k(n_k * (p + 1))$.

From a computational point of view, it is rather inefficient to calculate expressions like $\mathbf{Z}_k^T\mathbf{W}_k\mathbf{Z}_k$ and $\mathbf{Z}_k^T\mathbf{W}_k\mathbf{Y}_k$ as written, because this requires transposing a large matrix $\mathbf{Z}_k$. The above expressions can be calculated more efficiently by using weighted cross products of columns. Let us define a function $\mathbf{SSCP}(\mathbf{A}, \mathbf{W})$ of matrix $\mathbf{A}$ with arbitrary dimensions and diagonal matrix $\mathbf{W}$ of dimension $r * r$ (with $r$ the number of rows of $\mathbf{A}$) as follows:

$$\left(\mathbf{SSCP}(\mathbf{A}, \mathbf{W})\right)_{s,s'} = \sum_i w_i A_{i,s} A_{i,s'} = \sum_i w_i (\mathbf{A}^T)_{s,i} A_{i,s'} = (\mathbf{A}^T\mathbf{W}\mathbf{A})_{s,s'} \tag{4}$$

Here $s$ and $i$ are indices for a column and a row of matrix $\mathbf{A}$, respectively. The function $\mathbf{SSCP}$ (sum of squares and cross products) is similar to a covariance function except that one does not need to subtract the column mean before multiplying columns. In SAS, the SSCP matrix can be easily calculated using PROC CORR with option SSCP with input dataset $\mathbf{A}$ and weights for the $i_{th}$ individual $w_i$.

We can calculate matrices $\mathbf{Z}_k^T \mathbf{W}_k \mathbf{Z}_k$ and $\mathbf{Z}_k^T \mathbf{W}_k \mathbf{Y}_k$ by applying the **SSCP** function to a matrix that concatenates the columns of $\mathbf{Z}_k$ and $\mathbf{Y}_k$:

$$\boldsymbol{SSCP}(\mathbf{Z}_k \,||\, Y_k, \mathbf{W}_k) = \begin{pmatrix} \sum_i w_{i,k} \mathbf{Z}_{i,k}^T \mathbf{Z}_{i,k} & \sum_i w_{i,k} \mathbf{Z}_{i,k}^T Y_{i,k} \\ \sum_i w_{i,k} \mathbf{Z}_{i,k} Y_{i,k} & \sum_i w_{i,k} Y_{i,k}^2 \end{pmatrix} \quad (5)$$

Each $\boldsymbol{SSCP}(\mathbf{Z}_k \,||\, Y_k, \mathbf{W}_k)$ in Equation (5) can be easily computed at site $k$ from the individual-level data at that site. These highly summarized datasets can then be transferred to the analysis center to compute the combined SSCP dataset:

$$\boldsymbol{SSCP}(\mathbf{Z} \,||\, Y, \mathbf{W}) = \sum_k \boldsymbol{SSCP}(\mathbf{Z}_k \,||\, Y_k, \mathbf{W}_k) \quad (6)$$

The dataset in Equation (6) is created with the property TYPE explicitly set to SSCP (the property TYPE is a part of the SAS dataset metadata). This dataset can then be fed directly into the PROC REG procedure in lieu of an individual-level dataset to obtain the solution for Equation (3). Once the combined SSCP matrix is fed into PROC REG at the analysis center, the procedure automatically calculates many desired statistics. These include not only regression coefficient estimates $\widehat{\boldsymbol{\beta}}$, but also the variance estimate

$$\widehat{\sigma}^2 = \frac{1}{N-p} \sum_{k=1}^{K} \left[ \sum_{i=1}^{n_k} w_{i,k} \left( Y_{ik} - \widehat{\boldsymbol{\beta}}^T \mathbf{Z}_{i,k} \right)^2 \right], \quad (7)$$

inverse matrix $(\mathbf{Z}^T \mathbf{W} \mathbf{Z})^{-1}$ and the estimated covariance matrix

$$\widehat{cov}(\widehat{\boldsymbol{\beta}}) = \widehat{\sigma}^2 \, (\mathbf{Z}^T \mathbf{W} \mathbf{Z})^{-1} \quad (8)$$

along with collinearity diagnostics and a number of goodness-of-fit measures.

### 2.2.2 General iterative reweighted least squares algorithm

The above procedure is a special case of a more general IRLS algorithm for estimating the regression parameter $\boldsymbol{\beta}$ of a GLM. This general algorithm allows alternative choices of distribution, $\mu(\boldsymbol{\beta}^T \mathbf{Z}_{i,k})$ and $v(\boldsymbol{\beta}^T \mathbf{Z}_{i,k})$. For example, logistic regression is a GLM under a binomial outcome distribution with $\mu(\boldsymbol{\beta}^T \mathbf{Z}_{i,k}) = \frac{\exp(\boldsymbol{\beta}^T \mathbf{Z}_{i,k})}{1+\exp(\boldsymbol{\beta}^T \mathbf{Z}_{i,k})}$; $v(\boldsymbol{\beta}^T \mathbf{Z}_{i,k}) = \mu(\boldsymbol{\beta}^T \mathbf{Z}_{i,k})[1 - \mu(\boldsymbol{\beta}^T \mathbf{Z}_{i,k})]$. Poisson regression is another example under a Poisson outcome distribution with $\mu(\boldsymbol{\beta}^T \mathbf{Z}_{i,k}) = \exp(\boldsymbol{\beta}^T \mathbf{Z}_{i,k})$ and $v(\boldsymbol{\beta}^T \mathbf{Z}_{i,k}) = \mu(\boldsymbol{\beta}^T \mathbf{Z}_{i,k})$.

Unlike the special case of linear regression, IRLS for fitting a general GLM does not have an exact solution but iterates until convergence. Specifically, at each iteration $m + 1$ until a convergence criterion is met, IRLS solves:

$$\sum_{k=1}^{K} \sum_{i=1}^{n_k} \widetilde{w}_{i,k}(\widetilde{Y}_{i,k} - \boldsymbol{\beta}_{m+1}^T \mathbf{Z}_{i,k}) \mathbf{Z}_{i,k} = 0 \qquad (9)$$

for $\boldsymbol{\beta}_{m+1}$ where

$$\widetilde{w}_{i,k}(\boldsymbol{\beta}_m^T) \equiv w_{i,k} \mu'(\boldsymbol{\beta}_m^T \mathbf{Z}_{i,k}), \qquad (10)$$

$$\widetilde{Y}_{i,k}(\boldsymbol{\beta}_m^T) \equiv \frac{Y_{i,k} - \mu(\boldsymbol{\beta}_m^T \mathbf{Z}_{i,k})}{\mu'(\boldsymbol{\beta}_m^T \mathbf{Z}_{i,k})} + \boldsymbol{\beta}_m^T \mathbf{Z}_{i,k} \qquad (11)$$

and $\boldsymbol{\beta}_m$ representing the solution from the previous iteration $m$ (with $\boldsymbol{\beta}_0$ specified starting values). Both the redefined weight and outcome in Equations (10) and (11), respectively, change at each iteration, but the covariate vector $\mathbf{Z}_{i,k}^T$ remains the same. For the special case of linear regression, $\tilde{y}_{i,k} = Y_{i,k}$ and $\widetilde{w}_{i,k} = w_{i,k}$ and thus do not depend on $\boldsymbol{\beta}_m$. As expected, in this case the algorithm reduces to standard linear regression and does not require an iterative process.

In the more general case, the following describes a general implementation of IRLS to obtain an estimate of the regression coefficient $\boldsymbol{\beta}$ of a GLM using SSCP matrices and PROC REG in SAS. Following standard theory (McCullagh and Nelder 1989), the resulting estimate is a maximum likelihood estimator under distributional assumptions. This algorithm is implemented in the macro %*distributed_regression* that we describe in the next section.

1) For each iteration $m + 1$ at each site $k$, calculate the SSCP matrix

$$\boldsymbol{SSCP}(\mathbf{Z}_k \,||\, \widetilde{\mathbf{Y}}_{km}(\boldsymbol{\beta}_m), \widetilde{\mathbf{W}}_{km}(\boldsymbol{\beta}_m))$$

Bring these SSCP matrices from each site to the analysis center and calculate the combined SSCP matrix:

$$\boldsymbol{SSCP}(\mathbf{Z} \,||\, \widetilde{\mathbf{Y}}_m(\boldsymbol{\beta}_m), \widetilde{\mathbf{W}}_m(\boldsymbol{\beta}_m)) = \sum_k \boldsymbol{SSCP}(\mathbf{Z}_k \,||\, \widetilde{\mathbf{Y}}_{km}(\boldsymbol{\beta}_m), \widetilde{\mathbf{W}}_{km}(\boldsymbol{\beta}_m)) \qquad (12)$$

2) Feed the combined SSCP matrix from Equation (12) into PROC REG to solve for $\boldsymbol{\beta}_{m+1}$

3) Repeat until convergence is achieved. On the iteration $m + 1$ that meets the convergence criterion, $\widehat{\boldsymbol{\beta}} = \boldsymbol{\beta}_{m+1}$

After convergence is achieved, an additional iteration of Steps 1-3 will output the inverse of the matrix $\mathbf{Z}^T \widetilde{\mathbf{W}}(\widehat{\boldsymbol{\beta}}) \mathbf{Z}$. The extra iteration is necessary because at iteration $m + 1$ we do not know the matrix $\mathbf{Z}^T \widetilde{\mathbf{W}}(\boldsymbol{\beta}_{m+1}) \mathbf{Z}$. We only know the matrix $\mathbf{Z}^T \widetilde{\mathbf{W}}(\boldsymbol{\beta}_m) \mathbf{Z}$. Note that the weight does not depend on $\boldsymbol{\beta}$ and the extra step is not necessary for linear regression.

The covariance matrix can be calculated as:

$$\widehat{cov}(\widehat{\boldsymbol{\beta}}) = \mathbf{I}^{-1}(\widehat{\boldsymbol{\beta}}) = \phi \left[ \sum_{k=1}^{K} \sum_{i=1}^{n_k} w_{i,k} \mu'(\widehat{\boldsymbol{\beta}}^T \mathbf{Z}_{i,k}) \mathbf{Z}_{i,k} \mathbf{Z}_{i,k}^T \right]^{-1} = \phi \left( \mathbf{Z}^T \widetilde{\mathbf{W}}(\widehat{\boldsymbol{\beta}}) \mathbf{Z} \right)^{-1} \qquad (13)$$

where $\mathbf{I}(\boldsymbol{\beta}) = -\mathbf{H}(\boldsymbol{\beta}) = -\frac{\partial^2 l}{\partial \boldsymbol{\beta} \partial \boldsymbol{\beta}^T}$ is the negative of the Hessian matrix defined by:

$$\mathbf{I}(\boldsymbol{\beta}) = \sum_{k=1}^{K} \sum_{i=1}^{n_k} w_{i,k} \frac{\mu'(\boldsymbol{\beta}^T \mathbf{Z}_{i,k})}{\phi} \mathbf{Z}_{i,k} \mathbf{Z}_{i,k}^T \qquad (14)$$

The above expression for $\widehat{cov}(\widehat{\boldsymbol{\beta}})$ requires that the assumed probability distribution is correctly specified. The alternative sandwich variance estimator is robust to this assumption:

$$\widehat{cov}(\hat{\beta}) = \mathbf{I}^{-1} I_1 \mathbf{I}^{-1} \qquad (15)$$

where $\mathbf{I}(\boldsymbol{\beta})$ is as in Equation (14) and the matrix $I_1(\widehat{\boldsymbol{\beta}})$ can be calculated as:

$$I_1(\widehat{\boldsymbol{\beta}}) = \frac{N}{N-p} \sum_{k=1}^{K} \sum_{i=1}^{n_k} \frac{w_{i,k}^2 \left( Y_{i,k} - \mu(\widehat{\boldsymbol{\beta}}^T \mathbf{Z}_{i,k}) \right)^2}{\phi^2} \mathbf{Z}_{i,k} \mathbf{Z}_{i,k}^T \qquad (16)$$

The factor $\frac{N}{N-p}$ corresponds to the definition HC$_1$ for the robust estimator for linear regression in PROC REG. The expression can be evaluated at each site as $\boldsymbol{SSCP}(\mathbf{Z}_k, \mathbf{W_k^H})$ where the diagonal matrix of weights $\mathbf{W_k^H}$ has elements:

$$w_{i,k}^H = \frac{w_{i,k}^2 \left( Y_{i,k} - \mu(\widehat{\boldsymbol{\beta}}^T \mathbf{Z}_{i,k}) \right)^2}{\phi^2} \qquad (17)$$

After matrices $\boldsymbol{SSCP}(\mathbf{Z}_k, \mathbf{W_k^H})$ are brought to the analysis center, the matrix $I_1$ can be calculated as a sum of these matrices:

$$I_1 = \sum_k \boldsymbol{SSCP}(\mathbf{Z}_k, \mathbf{W_k^H}) \qquad (18)$$

Once the covariance matrix $\widehat{cov}(\widehat{\boldsymbol{\beta}})$ is calculated, the standard errors of $\widehat{\boldsymbol{\beta}}$ can be calculated by taking a square root of the corresponding diagonal elements of the matrix.

The key macros used by the IRLS algorithm are described in the **Appendix** A.

## 2.3 Distributed generalized linear model convergence criteria

We use the relative convergence criteria identical to the SAS relative convergence criteria specified by option XCONV. Let $\beta_s^m$ $m$ be the estimate of the parameter $s = 1, \ldots, p+1$ at iteration $m$. The regression criterion is satisfied if:

$$\max_s |\delta_s^{m+1}| < xconv\_value$$

where

$$\delta_s^{m+1} = \begin{cases} \beta_s^{m+1} - \beta_s^m, & |\beta_s^m| < 0.01 \\ \frac{\beta_s^{m+1} - \beta_s^m}{\beta_s^m}, & else \end{cases}$$

## 2.4 Goodness-of-fit measures

Our application also calculates a number of goodness-of-fit measures and tests for both linear and logistic DRA. Many of these measures can be evaluated exactly without individual-level data, because they can be expressed in terms of quantities that have the associative property, i.e., they can be added or multiplied regardless of how the numbers are grouped. These include the likelihood ratio for the global null hypothesis, $R^2$ (generalized $R^2$ for logistic regression), deviance, Akaike information criterion (AIC), corrected Akaike information criterion (AICC), and Bayesian information criterion (BIC). In **Appendix B** we give explicit expressions for these measures in the case of horizontally partitioned data.

For logistic regression, additional options are available to approximate the receive operating characteristic (ROC) curve, area under the ROC curve (AUC), and Hosmer-Lemeshow goodness-of-fit test when individual-level data cannot be shared. These measures depend on the overall ordering of data and cannot be calculated exactly without sharing some individual-level information. In Section 4.1, we describe our approximations of these statistics using pre-summarized data from data partners. The accuracy of these approximations depends on a choice of bin size used for data summarization; they approach the exact values as the bin size approaches one. Generally, the approximations should work well when the minimum bin size is chosen in accordance with federal, state, or institutional requirements or recommendations for privacy protection, e.g., *min_count_per_grp=6* as illustrated in Section 4.2.2.

## 3. How to use the DRA application for distributed linear and logistic regression

### 3.1 Overview

As an illustrative example of the DRA application, we describe a DRA in a DDN of four parties (

**Figure** 1) in this section. One party was designated as the semi-trusted third-party analysis center and the remaining three were data partners. Importantly, the DRA application follows a master-worker model design, where the analysis center directs the iterative DRA computations, while the data partners compute the required intermediate statistics. Below we describe the main steps involved in setting up and executing the DRA application.

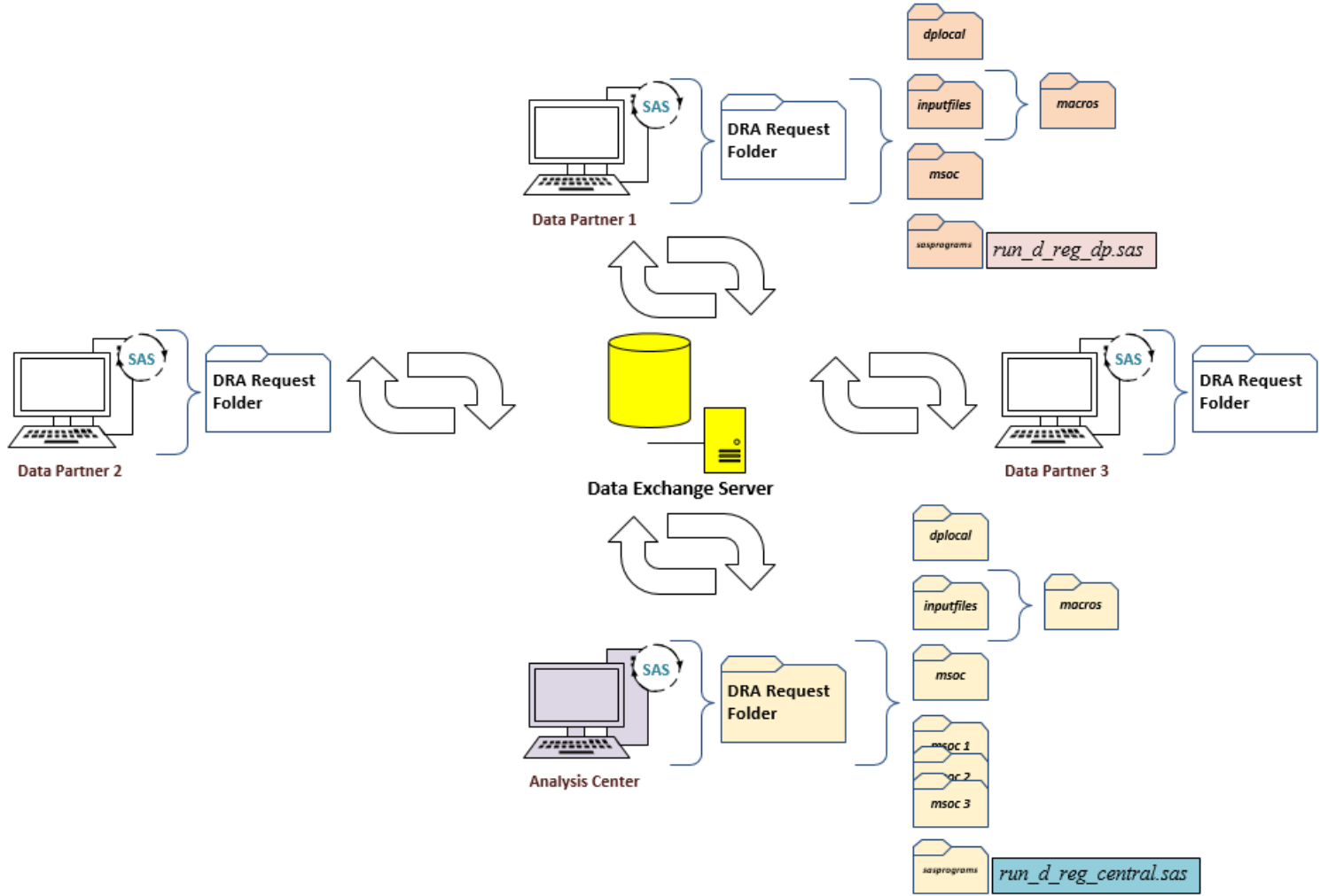

**Figure 1:** Distributed regression analysis (DRA) application with a semi-trusted third party (analysis center)

### 3.2 Creation of an individual-level analytic datasets at data partner sites

In this step, each data partner assembles an individual-level analytic dataset of the same structure. This can be done by executing a distributed program developed by the analysis center. Here we assume that this step was completed and the partitioned dataset described below represents the output of this first step.

#### 3.2.1    Example dataset

The publicly available "Boston housing dataset" is used to illustrate the steps involved in using the DRA application (Harrison and Rubinfeld 1978). The dataset included 506 observations of Boston medium housing prices and housing or neighborhood characteristics. Karr and colleagues used the dataset to illustrate the theoretical capability of conducting linear DRA in a horizontally partitioned data environment (Karr et al. 2004). To stay consistent with these authors, we also randomly partitioned the dataset into three data partners of sizes $n_1=172$, $n_2=182$, and $n_3=152$. Each dataset included the following continuous variables – housing price ("medv"), crime per capita ("crim"), industrialization ("indus"), and distance to employment centers ("dis"). Housing price served as the dependent variable for the linear DRA. For logistic DRA, we dichotomized housing price into "low" or "high" (below or above the median) and used the derived binary variable ("medv_high_flag") as the dependent variable. The independent variables in both models included crim,  indu, and dis, and indicator variables for data partner sites (dummy_dp_var2, dummy_dp_var3).

#### 3.2.2    Directory structures required to execute the DRA application

There are two separate SAS packages in our DRA implementation, one for the analysis center and one for the data partners. Each package employs directory structures used by the Sentinel System **(Figure 1)** (Sentinel System 2017; Her et al. 2018). At the data partners, this directory structure is composed of four subdirectories: *sasprogram, inputfiles, dplocal,* and *msoc*. At the analysis center, the directory structure has additional *msoc&dp_cd* subdirectories that receive data from corresponding data partners (*&dp_cd* corresponds to a unique data partner ID; see **Error! Reference source not found.**). The subdirectory *sasprogram* contains the master wrapper program: *run_d_reg_dp_templ.sas* at the data partner site and the master wrapper program *run_d_reg_central_templ.sas* at the analysis center. **Table 1** summarizes the required directory structure used to organize the DRA application.

| Directory | Data Partners | Analysis Center |
|---|---|---|
| *dplocal* | Files to be kept at a data partner (e.g. final individual-level regression dataset with all initial variables plus predicted outcome, residuals, etc) | Intermediate permanent files (e.g., combined SSCP matrix from all data partners) |
| *sasprogram* | *run_d_reg_dp_templ.sas* | *run_d_reg_central_templ.sas* |
| *inputfiles* | Files received from the analysis center (e.g., updated regression parameters). Includes a subdirectory for macros. | Initial input files and files to be sent to the data partners (e.g., initial and updated regression parameters). Includes a subdirectory for macros. |
| *msoc* | Files to be shared with the analysis center (e.g., contribution to the SSCP matrix from a data partner) | Final distributed regression analysis results |
| *msoc&dp_cd* | N/A | Subdirectories with returned summary-level data from data partners. |

**Table 1:** Distributed regression analysis (DRA) application common directory structure

### 3.2.3 SAS package for data partners

To simplify the DRA process for data partners, the analysis center prepares a package as a zip file with the whole directory structure. All necessary macros are in the *macros* subdirectory, and the main SAS wrapper is located in the subdirectory *sasprograms*. Once the package is received by a data partner, it is unzipped into a pre-defined root directory to create the required directory structure in **Figure 1**. After this is done, the user at the data partner site only needs to update a few site-specific parameters (discussed below). All other site-specific parameters such as the file path to the subdirectories (*inputfiles, msoc, dplocal, macros*, etc) and the corresponding SAS libraries are derived within the wrapper template. The wrapper also has code to compile (*%include*) all necessary macros and calls *%dp_main*, the main macro to be executed by data partners that also calls other macros as necessary. This main macro has only one parameter – a data partner identifier called *dp_cd*. No regression-related parameters are specified at the data partner site. Instead, all parameters are specified in the main macro %***distributed_regression*** at the analysis center and passed to data partners in the form of a SAS dataset, *vars_nm_value_pairs*, which stores pairs of parameters names and values (see **Table 2**). Below we focus on the parameters that have to be updated by users at a given data partner site and give an outline of the main wrapper structure. In the example below, the root directory for all DRA requests at the data partner with *dp_cd=1* is called \*distrib_regr_all_requests_dp1\* , the request id is called *request_1* and the full path for the request is \*distrib_regr_all_requests_dp1\ request_1\*.

**Main wrapper template for the data partner package,** *run_d_reg_dp_templ.sas*

```
/* Edit data partner numeric code dp_cd using a lookup table for data partners. Maximum length
3 digits.*/

%LET dp_cd=1;

 * Edit root directory for requests related to distributed regression;
 * This is the directory into which you unzipped the zip file with the current request;

%LET sites_prg_root_dp=\distrib_regr_all_requests_dp1\;

/* Edit directory which has the input regression dataset.
   In production it is likely to point to the full path of the sub-directory "dplocal" for the initial
request which created the analytic dataset.   */

  %LET data_in_dir=&sites_prg_root_dp.MSReqID_for_step0\dplocal\;

/* Edit parameter min_count_per_grp to define minimum count per cell for summarized data
returned to analysis center.   This affects only datasets used for residual analysis and some
goodness-of-fit measures (ROC and Hosmer-Lemeshow statistic for logistic regression).
  It has no effect on the regression coefficients or any other statistics */

 %LET min_count_per_grp=6;
    *******                     END OF USER INPUT                      ******;

  /*The following are derived from the parameters entered above*/
/*The MSReqID is defined at the analysis center before distributing request to all data partners.
  The MSReqID should be the same name as the directory for the request.
  All data partners participating in the request should have the same MSReqID.;
*/

%LET MSReqID = request_1;

/* Derive full path for standard Sentinel folders */

%let DPLOCAL = &sites_prg_root_dp.&MSReqID./dplocal/ ;
%let MSOC = &sites_prg_root_dp.&MSReqID./msoc/ ;
%let INFOLDER = &sites_prg_root_dp.&MSReqID./inputfiles/ ;
%let SASPROGRAMS = &sites_prg_root_dp.&MSReqID./sasprograms/ ;
%let SASMACR=&Infolder.macros/;

/*   Other statements which include: required SAS options, assignments for SAS libraries and
some additional macro variables and the statements which include (compile) all macros in the
data partner package. */
………………
```

```
* Call to main data partner macro. The macro calls all other macros as needed;

%dp_main(dp_cd=&dp_cd.);
/*   End main wrapper for the data partner package */
```

### 3.2.4 SAS package for analysis center

The main macro in the analysis center package is called %*distributed_regression*. The *run_d_reg_central_templ.sas* wrapper is used to define site-specific macro parameters (e.g., root directory for the package), compile *(%include)* necessary macros, and call the main macro %*distributed_regression*. It is also used to specify all regression-related parameters, including the dependent and independent variables, convergence criteria, the list of participating data partners, and type of regression (e.g., linear). The complete list of user-specified parameters is described in **Appendix C.** Currently, the macro %*distributed_regression* performs linear, logistic, and Cox DRA. This paper is focused on linear and logistic DRA, which are particular cases of GLMs. Cox DRA is implemented via a different algorithm and is discussed in our companion paper.

The wrapper for the SAS package at the analysis center has similar functionality as the wrapper at the data partner sites, but it defines additional subdirectories and SAS libraries *msoc&dp_cd1 – msoc&dp_cdN* to store the returned summary-level data from data partners. The templates for SAS wrappers *run_d_reg_central_tmpl.sas* and *run_d_reg_dp_templ.sas* can be downloaded together with all the macros from https://www.sentinelinitiative.org/sentinel/methods/utilizing-data-various-data-partners-distributed-manner. Below we focus on the parameters that have to be updated by users at the analysis center and give an outline of the main wrappers structure. We describe how to use the main macro %*distributed_regression* in Section 3.6. In the example, below the root directory for all DRA requests at the at the analysis center is called \*distrib_regr_all_requests_central\*, the request id is called *request_1* and the full path for the request directory is \ *distrib_regr_all_requests_central* \ *request_1*\

**Main wrapper template for the analysis center package,** *run_d_reg_central_templ.sas*

```
/* The parameter dp_cd (data partner code) for the analysis center should be always set to 0.*/

%LET dp_cd=0;

/* Specify list of data partner codes participating in distributed regression*/
    %LET dp_cd_list=1 2 3;

 * Edit root directory for all requests related for distributed regression at the analysis center. ;
%LET sites_prg_root_dp=\distrib_regr_all_requests_central\;

/*The MSReqID is defined at the analysis center.
  Should be the same as name of the subdirectory for the request.
  Should have the same value as the one pre-specified in wrappers for data partners
  participating in the same request.;
```

```
*/

%LET MSReqID = request_1;

/* Derive full path for standard Sentinel folders */
%let DPLOCAL = &sites_prg_root_dp.&MSReqID./dplocal/ ;
%let MSOC = &sites_prg_root_dp.&MSReqID./msoc/ ;
%let INFOLDER = &sites_prg_root_dp.&MSReqID./inputfiles/ ;
%let SASPROGRAMS = &sites_prg_root_dp.&MSReqID./sasprograms/ ;
%let SASMACR=&Infolder.macros/;

/*    Other statements which include: required SAS options, assignments for SAS libraries and
some additional macro variables and the statements which include (compile) all macros in the in
the package for the analysis center (AC). */
………………

* Call main macro %*distributed_regression*. The macro calls all other macros as needed;
```

### 3.3 Execution of SAS programs

After necessary parameters are updated in the SAS wrappers, users at data partners and the analysis center can start the execution of their SAS programs at any time within a mutually agreed time window. All programs run continuously but the program at the data partner sites goes into a waiting mode immediately. The program constantly checks the *inputfiles* subdirectory for the arrival of input files from the analysis center. More specifically, it checks for the existence of the trigger file *files_done.ok* which is always the last file created at any data transfer step. At the analysis center, the program creates a set of files in its *inputfiles* subdirectory which are picked up by the data transferring software (e.g., PopMedNet, see Section 3.4) and transferred to the data partners via a data exchange server. These files include the parameters dataset *vars_nm_value_pairs* (see **Table 2**) and files required by the DRA algorithm (e.g., a dataset with initial regression estimates).

The process then continues using a data exchange mechanism described in Section 3.4 until the program at the analysis center issues a stop instruction to the data partners. This occurs when either the regression algorithm converges (first iteration for linear), reaches the pre-specified maximum number of iterations, or the program catches an error in the process. To stop all SAS processes, the central program sets parameters *last_iter_in* and *end_job_dp_in* to 1. These parameters are passed to the data partners' SAS programs using the parameters dataset *vars_nm_value_pairs* (**Table 2**). At the data partners, the condition *last_iter_in=1* and *end_job_dp_in =1* instructs the program to calculate data for final regression statistics (goodness-of-fit measures, summary of residuals, etc.), create the empty file *job_done.ok* in the *msoc* subdirectory, and exit the data partner's SAS program. Finally, after the last batch of files from all data partners are downloaded to the analysis center, the SAS program at the analysis center performs final calculations, creates the empty file *job_done.ok* in the *inputfiles* subdirectory, and exits SAS. This concludes the DRA process. If the program at a data partner or analysis center catches an error, it creates the file *job_fail.ok* instead of file *job_done.ok*.

| M_var_nm | M_var_value |
|---|---|
| reg_ds_in | linear_karr_2005 |
| independent_vars | crim indus dis dummy_dp_var2 dummy_dp_var3 |
| dependent_vars | medv_high_flag |
| regr_type_cd | 2 |
| iter_nb | 1 |
| last_iter_in | 0 |
| end_job_dp_in | 0 |

**Table 2:** Sample records from the parameters dataset *vars_nm_value_pairs*. Column M_var_nm contains the name of the macro variable, and the column M_var_value contains its value.

### 3.4 Data transfer mechanism

Data transfer in DDNs is challenging because all computers are located behind their respective firewalls. One way to solve this problem is to use an intermediate data exchange server that allows all sites to access, upload, and download files to and from that server. We adapted this approach in our implementation of the DRA.

In this section, we describe the minimal requirements to integrate our DRA application with a data transfer software. The steps below describe what files the SAS process expects from the data transferring software and what files the data transferring software should expect from SAS. These steps are repeated at each iteration.

1. The SAS program at the analysis center outputs files into its subdirectory *inputfiles*. At the end of outputting, SAS creates two additional files that facilitate data transfer: an empty trigger file *files_done.ok* and a manifest file *file_list.csv*. The first file informs the data transferring software that files are ready to be transferred from the analysis center to the data partners, and the second file lists the files to be transferred.

2. The SAS program at each data partner site monitors its subdirectory *inputfiles* for the appearance of the trigger file *files_done.ok*. This file must be created by the data transferring software after it transferred all datasets from the analysis center to the data partner. Once the SAS program finds the trigger file *files_done.ok*, it deletes the trigger file, resumes execution, and calculates intermediate statistics. The SAS program also creates the trigger file *files_done.ok* and the manifest file *file_list.csv* in the subdirectory *msoc*. The trigger file informs the data transferring software that files are ready to be transferred from the data partner to the analysis center, and the manifest file lists the files that has to be transferred.

3. The SAS program at the analysis center monitors its subdirectories *msoc&dp_cd1*, *msoc&dp_cd2, ...msoc&dp_cdN* for the appearance of the trigger file *files_done.ok*. The trigger file in each of these subdirectories must be created by the data transferring software after it transferred all datasets from the corresponding data partners. Once SAS finds the trigger file *files_done.ok* in all of the above *msoc&dp_cdN* subdirectories, it resumes execution and calculates updated regression parameter estimates.

4. The process repeats until the SAS program at the analysis center issues instruction to stop execution to all SAS processes using mechanism described in the Section 3.3.

In the next section, we describe automated, semi-automated and manual implementation of our DRA application via PopMedNet, the data transferring software used by several DDNs.

### 3.4.1 Automated data transfer via PopMedNet

We have successfully integrated the DRA application with PopMedNet, an open-source query distribution software application that supports automatable file transfer between an analysis center and data partners. Detailed description of the file transfer process is available elsewhere (Her et al. 2018) and we briefly summarize it here. The PopMedNet DataMart Client is a Windows® application installed at the analysis center and all participating data partners. It monitors the subdirectories described above for the presence of the trigger file *files_done.ok*, reads the manifest file, and transfers data found in the manifest file to the other side (analysis center to data partners, and vice versa) using the PopMedNet server. It also creates the trigger file *files_done.ok* in the corresponding folder once the data transfer is completed.

### 3.4.2 Semi-automated data transfer via PopMedNet

In this scenario the process proceeds in a similar fashion as in the automated mode, but the user is prompt to manually upload the files to the PopMedNet DataMart Client for transfer. This allows the data partner to inspect the files before uploading and transferring them to the analysis center. We introduce this mode primarily to facilitate adoption of the DRA application because it allows data partners to see the highly summarized datasets that are being transferred at each step. We do not recommend using this mode if participating sites have already established collaborative relationships or agreements to share the summary-level data.

### 3.4.3 Manual data transfer

As the name suggests, in this mode, a user performs all functions manually. For example, the user at the analysis center monitors its subdirectory *inputfiles* for the appearance of the trigger file *files_done.ok* and then manually uploads the files listed in the manifest file to the data exchange server. The user also needs to manually delete the trigger file in the subdirectory *inputfiles* to ensure its appearance in the next iteration prompts another manual transfer process. The user at the data partner also has to manually monitor the PopMedNet DataMart Client for a status change and then download the files listed in the manifest file to its subdirectory *inputfiles*. After all files are downloaded, the user must deposit the trigger file *files_done.ok* to its *inputfiles* subdirectory to resume SAS execution with the new files. This is a tedious process and should be only used when the automated or semi-automated options are not available.

**3.5 Testing the DRA application without data transferring software**

The macro %*distributed_regression* has a test mode that can perform DRA on a single computer without a data transferring software. It allows researchers to test most of the SAS code, including all code for statistical calculations, most of the code for error handling, and some elements of the code necessary for data transfer.

To run the macro in the test mode, the parameter *test_env_cd* has to be set to 1 (see example below). The directory structure required for this mode is the same as the directory structure for the analysis center in the production mode described above. The template for the SAS wrapper (*run_d_reg_test_mode_tmpl.sas*) is similar to the wrapper for the analysis center, with the exception that an *%include* statement compiles the macros from both the analysis center and data partners. It also defines the subdirectory *&data_in*, which contains the datasets for the "data partners". The datasets corresponding to each data partner should be placed in the subdirectory *&data_in* and should be named using the naming convention: *®_ds_in._&dp_cd* (e.g., LINEAR_KARR_2005_1, LINEAR_KARR_2005_2). The test mode wrapper can be downloaded as a part of the analysis center package (Sentinel System 2018). Below we give an example and outline how the test mode works without data transferring software.

Example of the call to the macro %*distributed_regression* in the test mode:

```
%distributed_regression(RunID=dl16
                ,dp_cd_list=1 2 3
                ,reg_ds_in=LINEAR_KARR_2005
                ,dependent_vars=medv_high_flag
                ,independent_vars= crim indus dis dummy_dp_var2
dummy_dp_var3                        ,regr_type_cd=2
                ,tbl_intial_est=Model_Coeff_0
                ,test_env_cd=1);
```

The program runs continuously in a single SAS session. As in production mode, the final regression results can be found in the subdirectory *msoc*.

Here we provide a brief outline of how the test mode works. When parameter *test_env_cd=1*, the macro %*distributed_regression* calls the macro *%_t_loop_through_dp*. which is executed between the code that creates the files to be sent to the data partners (macro %*d_reg_central_step1*) and the code that receives files from the data partners (macro %*d_reg_central_step2*). See code snippet below:

```
%d_reg_central_step1;
/*Execute distributed regression code at each data partner. Test mode only */

%IF  &test_env_cd. NE 0 %THEN %DO;
       %_t_loop_through_dp(prefix=&prefix., dp_cd_list=&dp_cd_list.);
%END;
```

```
/* Analysis center step 2*/
%LET prev_reg_conv_in=®_conv_in.;
%d_reg_central_step2;
```

The macro *%_t_loop_through_dp* loops through the list of "data partners" codes specified in the parameter *dp_cd_list* and for each *&dp_cd* calls the main macro in the data partner package, *%dp_main*. This macro defines *&msoc_dir,* which in production mode is defined in the data partner wrapper and points to the subdirectory *msoc*. In the test mode, *&msoc_dir* is define in the macro *%_t_loop_through_dp* and points to the subdirectory *msoc&dp_cd* for data partner *&dp_cd*. For example, the *&msoc_dir* points to the subdirectory *msoc1* when *&dp_cd=1*. As a result, the output files from the macro *%dp_main* for *&dp_cd=1* are written to the subdirectory *msoc1*, which is where the program from the analysis center expects to find them. This eliminates the need for a data transferring software in the test mode.

### 3.6 Examples of using the main macro

In this section, we show some examples of using %*distributed_regression*, the main macro at the analysis center. The parameters explained below should be sufficient for most practical applications. The complete list of all parameters and their descriptions can be found in **Appendix C**.

*Example 1.* The code below runs distributed linear regression on the "Boston housing dataset" described in Section 3.2.1.

```
/*
 Parameter RunID specifies an identifier for a given macro call. It is used to form a prefix
%let prefix=&RunID. for all output datasets names.
 Parameter dp_cd_list specifies a list of data partner sites participating in the current request.
 Parameter reg_ds_in specifies the name of the input dataset for regression at a data partner
site, the name and structure the same at all sites. The dataset must be located in the SAS
library data_in defined in the data partner wrapper.
 Parameters dependent_vars and independent_vars specify dependent and independent
variables for the regression.
 Parameter regr_type_cd defines the type of the regression: 1- linear; 2- logistic;
*/

 %distributed_regression(RunID=dr1
                    ,dp_cd_list=1 2 3
                    ,reg_ds_in=LINEAR_KARR_2005
                    ,dependent_vars=medv
                    ,independent_vars=crim indus dis dummy_dp_var2 dummy_dp_var3
                    ,regr_type_cd=1
                    ) ;
```

***Example 2.*** This example runs distributed logistic regression on the same dataset and the same set of independent variables as in Example 1. It specifies two optional parameters *tbl_inital_est* and *xconv*.

```
/* Parameter tbl_intial_est names the table with initial guesses (starting values) for the
regression parameter estimates. Must have a column for each of the parameter estimates which
has the same name as the corresponding covariate. It should be located in the SAS library
named infolder. In the example below dataset Model_Coeff_0 has all initial values equal to 0
except for the intercept which is set to average of the outcome variable. This is the same set of
the initial values that is used by default by PROC LOGISTIC on a combined dataset.

Parameter xconv specifies relative convergence criteria.
*/

 %distributed_regression(RunID=dl16
                ,dp_cd_list=1 2 3
                ,reg_ds_in=LINEAR_KARR_2005
                ,dependent_vars=medv_high_flag
                ,independent_vars= crim indus dis dummy_dp_var2 dummy_dp_var3
                ,regr_type_cd=2
                ,tbl_intial_est=Model_Coeff_0
                ,xconv=1e-4) ;
```

### 3.7 Creation of output tables

The macro %*distributed_regression* creates final output datasets in the subdirectory *msoc* at the analysis center. All datasets from a given execution of the macro have the same prefix determined by the parameter *RunID*. The structure of most of these datasets was modeled after their corresponding datasets generated by PROC REG (linear) and PROC LOGISTIC/GENMOD (logistic). The complete list of output datasets and their description is given in **Appendix** D. The output can be generated by printing the output tables in the subdirectory *msoc,* or by using macro %*create_greg_rpt* included with the package at the analysis center. An example of how to use this macro is shown in the wrapper template *run_d_reg_central_tmpl.sas*.

### 3.8 Operating systems in which the DRA application can be used

The SAS packages for both the analysis center and data partners can be executed on any operating system on which SAS can be installed. These include Windows, UNIX, and Linux and some others. The PopMedNet DataMart Client is a Windows application. For this reason, most of our testing was done on Windows machines. However, we were also able to successfully test the DRA application with a data partner on a Linux server, where we placed the SAS package on a Linux server directory accessible to a Windows network as a mapped drive. This allowed the PopMedNet DataMart Client to access the same file system as the SAS program.

## 4. Example output created by macro %*distributed_regression*

We tested our DRA application on several datasets, including two publicly available datasets, a simulated data, and empirical datasets from three data partners in the Sentinel System. The examples of the full report generated by the macro %*create_greg_rpt* can be found online at https://www.sentinelinitiative.org/sentinel/methods/utilizing-data-various-data-partners-distributed-manner.

### 4.1 Main results from distributed linear and logistic regression

In this section, we report the parameter estimates, standard errors, and some goodness-of-fit-measures produced by the main macro %*distributed_regression* on the Boston housing dataset, one of the publicly available datasets used in our development of the DRA application (**Table 3 to Table 6**).

| Statistic | Value |
|---|---|
| Root MSE | 7.475420 |
| Dependent Mean | 22.532806 |
| Coeff Var | 33.175717 |
| R-Square | 0.345895 |
| Adj R-Sq | 0.339354 |
| _AIC_ Akaike's information criterion | 2041.723909 |
| _BIC_ Sawa's Bayesian information criterion | 2043.867621 |
| _SBC_ Schwarz's Bayesian criterion | 2067.083129 |

**Table 3:** Fit statistics for distributed linear regression from Example 1 described in Section 3.6.

| Variable | DF | Parameter Estimate | Standard Error | P-Value | Lower 95% CL Parameter | Upper 95% CL Parameter | Heteroscedasticity Consistent Standard Error |
|---|---|---|---|---|---|---|---|
| Intercept | 1 | 31.79302 | 1.68240 | <.0001 | 28.48757 | 35.09847 | 1.55065 |
| crim | 1 | -0.23283 | 0.04755 | <.0001 | -0.32626 | -0.13940 | 0.04661 |
| indus | 1 | -0.51302 | 0.08165 | <.0001 | -0.67343 | -0.35260 | 0.07754 |
| dis | 1 | -1.05423 | 0.22632 | <.0001 | -1.49888 | -0.60957 | 0.21689 |
| dummy_dp_var2 | 1 | 4.62054 | 0.88611 | <.0001 | 2.87958 | 6.36150 | 0.76374 |
| dummy_dp_var3 | 1 | -1.22053 | 1.04369 | 0.2428 | -3.27109 | 0.83003 | 1.09139 |

| Variable | Heteroscedasticity Consistent P-Value | Heteroscedasticity Consistent Lower 95% CL Parameter | Heteroscedasticity Consistent Upper 95% CL Parameter |
|---|---|---|---|
| Intercept | <.0001 | 28.74642 | 34.83962 |
| crim | <.0001 | -0.32440 | -0.14125 |
| indus | <.0001 | -0.66537 | -0.36066 |
| dis | <.0001 | -1.48036 | -0.62809 |
| dummy_dp_var2 | <.0001 | 3.12002 | 6.12107 |
| dummy_dp_var3 | 0.2640 | -3.36481 | 0.92375 |

**Table 4:** Parameter estimates for distributed linear regression from Example 1 described in Section 3.6.

|                         | *Fit Statistics* |
|-------------------------|-----------------:|
| *Criterion*             | *Value*          |
| Log Likelihood          | -261.03195       |
| Full Log Likelihood     | -261.03195       |
| AIC (smaller is better) | 534.06390        |
| AICC (smaller is better)| 534.23223        |
| BIC (smaller is better) | 559.42312        |
| R-Square                | 0.29797          |
| Max-rescaled R-Square   | 0.39740          |

**Table 5:** Fit statistics for distributed logistic regression from Example 2 described in Section 3.6.

| Variable | DF | Parameter Estimate | Standard Error | P-Value | Lower 95% CL Parameter | Upper 95% CL Parameter | Robust Standard Error |
|---|---|---|---|---|---|---|---|
| Intercept | 1 | 1.68778 | 0.53174 | 0.0015033 | 0.64558 | 2.72998 | 0.49189 |
| crim | 1 | -0.15315 | 0.04653 | 0.0009974 | -0.24435 | -0.06195 | 0.04258 |
| indus | 1 | -0.10329 | 0.02570 | 0.0000583 | -0.15366 | -0.05292 | 0.02383 |
| dis | 1 | -0.16344 | 0.07341 | 0.0259855 | -0.30732 | -0.01956 | 0.07045 |
| dummy_dp_var2 | 1 | 1.33919 | 0.27156 | 8.1622E-7 | 0.80694 | 1.87144 | 0.26679 |
| dummy_dp_var3 | 1 | 0.31595 | 0.37325 | 0.3972768 | -0.41560 | 1.04750 | 0.38528 |

| Variable | Robust P-Value | Robust Lower 95% CL Parameter | Robust Upper 95% CL Parameter |
|---|---|---|---|
| Intercept | 0.0006 | 0.72370 | 2.65186 |
| crim | 0.0003 | -0.23660 | -0.06970 |
| indus | <.0001 | -0.14999 | -0.05659 |
| dis | 0.0203 | -0.30152 | -0.02536 |
| dummy_dp_var2 | <.0001 | 0.81629 | 1.86209 |
| dummy_dp_var3 | 0.4122 | -0.43919 | 1.07109 |

**Table 6:** Parameter estimates for distributed logistic regression from Example 2 described in Section 3.6.

**4.2 Model diagnostics in distributed regression**

**4.2.1   Residual diagnostics**

Due to privacy considerations, residual analysis in the case of DRA requires a different approach than standard regression analysis. To comply with the privacy requirements of data partners, our approach leaves the final individual-level dataset with residuals and predicted values at the data partner sites and brings back only summarized results to the analysis center. Each data partner has an option to define a minimum number of records that must be summarized (minimum number per cell) by specifying the macro parameter *min_count_per_grp* in their master wrapper program (see template *run_d_reg_dp_templ.sas*). If the parameter *min_count_per_grp* is not specified by a data partner then the parameter *min_count_per_grp_glob*, specified in the macro *%distributed_regression* is used. In the examples below, we used *min_count_per_grp=6*.

To summarize the residual analysis, we first sort the individual-level output dataset at each data partner site by the predicted value of the outcome $\mu$. We then group the data into bins based on percentiles of $\mu$ and calculate means of the predicted and actual outcomes, residuals, and other variables for each bin. The number of observations can vary slightly between bins due to ties (see **Appendix** F for grouping algorithm). The number of bins in the summary dataset created at the data partner sites can be modified by changing the parameter *groups* in the main macro *%distributed_regression*. However, if the value of this parameter is too large to satisfy the constraint set by *min_count_per_grp*, the actual number of bins is decreased accordingly by the program. The summarized datasets from each data partner are brought to the analysis center and combined into a single dataset, which can then be used to visually evaluate the goodness-of-fit (see description of the dataset *&prefix0.resid_sum_by_pct* in **Appendix** D).

In Figure 2,**Error! Reference source not found.** we illustrate a plot of the mean observed response vs. the mean predicted response for logistic DRA by bin. In this example, we used 10 groups (deciles) at each data partner. Random scatter of data points around the diagonal reference line suggests reasonable model fit. The graph is also useful in assessing differences between data partners. For example, the plot indicates that the probability of the outcome is systematically higher for data partner 2 versus 3 and 1.

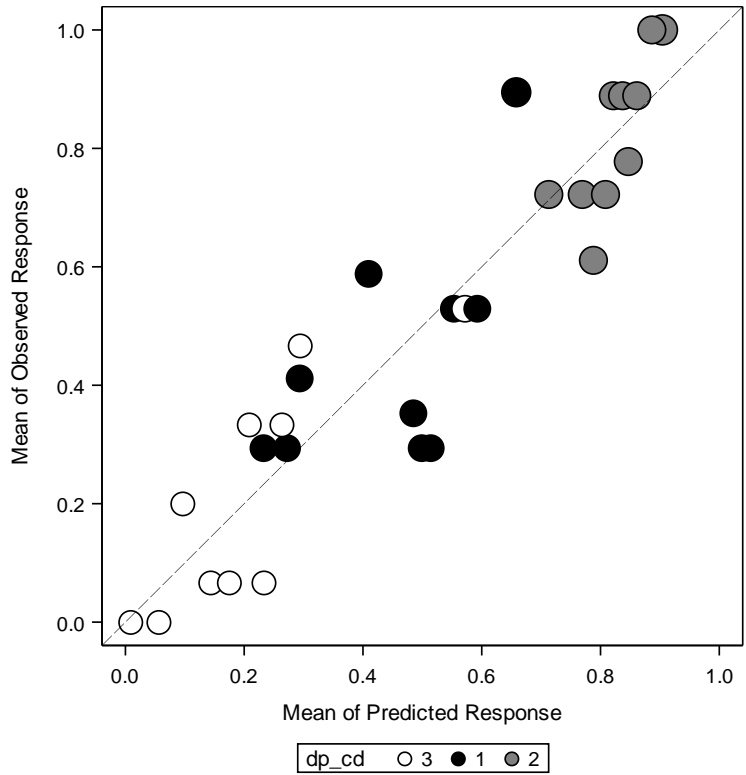

**Figure 2.** Mean observed response vs. mean predicted response for distributed logistic regression from Example 2 described in Section 3.6. The data are grouped into bins by decile of predicted values. The radius of each bubble is proportional to the number of observations in a bin. Different symbols represent data points from different data partners. If the model is correctly specified, the data points are expected to scatter randomly around the diagonal line.

### 4.2.2 Approximations for calculating ROC curves and the Hosmer-Lemeshow statistic

The calculation of the ROC curve, AUC, and Hosmer-Lemeshow statistic requires ordering by predicted value $\mu$. To calculate these quantities exactly one needs to bring back to the analysis center all distinct values of $\mu$ from the data partners. Such an approach was previously adopted by Wu et al (Wu et al. 2012). However, bringing back individual-level predicted values presents an additional privacy risk. In our experience, many data partners have policies requiring that any individual-level data be summarized before they can be shared. Also, some predicted values may be unique to an individual, particularly when covariates include continuous variables with rare extreme values or a few binary variables. In **Appendix** E, we describe an approximate approach that uses a pre-summarized dataset from each data partner. It allows each data partner to control the level of summarization via the parameter *min_count_per_grp*. When this parameter is set to *min_count_per_grp=1* at all data partners, our approach recovers the ROC curve, AUC, and Hosmer-Lemeshow statistic computed from individual-level data.

The logistic DRA results for the ROC curve and AUC with *min_count_per_grp=6* are shown in **Figure 3**. For comparison, we also provided the ROC curve obtained using PROC LOGISTIC on the individual-level dataset in the same figure. As expected, the ROC curve obtained using individual-level data provides more detail (there are more data points) but is generally similar to the curves produced by our approximation. Our approximation for AUC differs only by 0.1% from the exact results based on individual-level data. We also tested our approximation using different datasets and values of *min_count_per_grp*. As expected, the smaller the value of *min_count_per_grp* $/n_k$ the more accurate the approximation. For example, the AUC difference was only 0.01% for datasets with about 5,000 observations with *min_count_per_grp=6*. We also confirmed that we reproduced the exact result whenever we set *min_count_per_grp=1*.

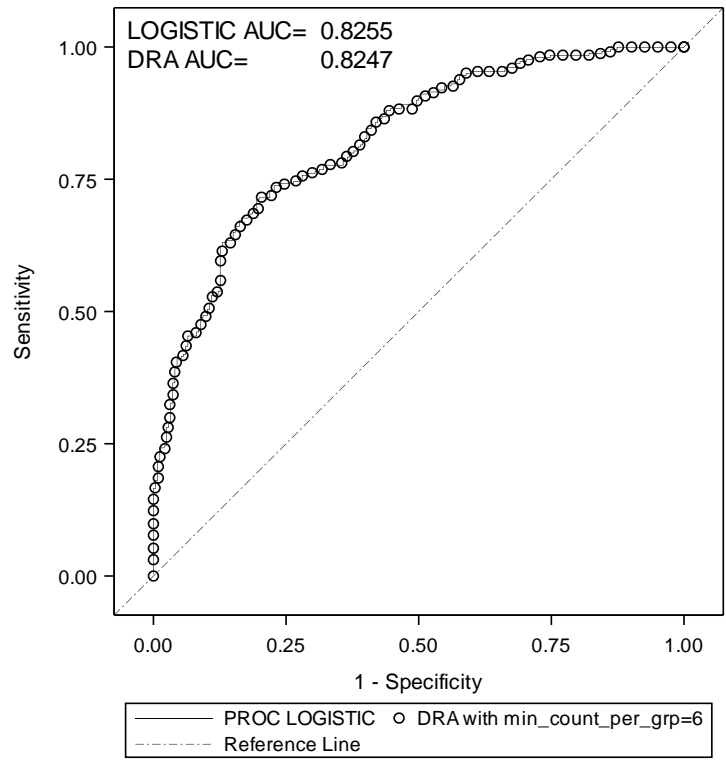

**Figure 3:** Comparison of the distributed logistic regression approximation for the receiver operating characteristic (ROC) curve with pooled individual-level logistic regression from Example 2 described in Section 3.6.

The logistic DRA results of our approximation for the Hosmer-Lemeshow statistic are shown in

**Table 7**. For comparison, we provide results obtained using PROC LOGISTIC on the combined individual-level dataset in the same table. As expected, the DRA results for *min_count_per_grp=1* is the same as those obtained from PROC LOGISTIC. The value of the test statistic for DRA with *min_count_per_grp=6* is empirically close to the value obtained from PROC LOGISTIC with individual-level data. However, at the 0.05 statistical significance level, conclusions regarding statistical significance would be different given a slightly larger p-value in the DRA case. As pointed out previously, the Hosmer-Lemeshow test must be interpreted with great caution, especially when the p-value is close to the cutoff level (Allison 2018).

| | *Hosmer and Lemeshow Goodness-of-Fit Test* | | | |
|---|---|---|---|---|
| | *Chi-Square* | *DF* | *ValueDF* | *Pr > ChiSq* |
| DRA with min_count_per_grp=6 | 15.43980 | 8 | 1.9299754 | 0.0511 |
| DRA with min_count_per_grp=1 | 15.87051 | 8 | 1.9838143 | 0.0443 |
| PROC LOGISTIC on combined dataset | 15.87051 | 8 | 1.9838143 | 0.0443 |

**Table 7:** Comparison of the distributed logistic regression approximation for the Hosmer-Lemeshow test with pooled individual-level logistic regression from Example 2 described in Section 3.6.

### 4.3 Comparison of the results for distributed regression with results obtained using standard SAS procedures on combined individual-level dataset

In all our tests using publicly available datasets, a simulated dataset, and empric data from three Sentinel data partners, our DRA algorithms produced regression parameters and standard errors estimates that are in complete agreement with the results produced by standard SAS procedures on the combined data. Specifically, they were within machine precision (1E-16) when we used the same input parameters, including initial guesses for parameter estimates.

In

**Table** 8 and

**Table** 9, we compare results DRA of the "Boston housing dataset" versus standard SAS procedures applied to a combined individual-level dataset. For linear regression, the results based on the combined individual-level dataset were obtained using PROC REG. For logistic regression, the results, except for robust standard errors, were obtained using PROC LOGISTIC. The results for robust standard errors, were obtained using PROC GENMOD with a REPEATED statement and defining a dummy cluster variable with one observation (individual) per cluster. In this case, the covariance matrix obtained from PROC GENMOD is the same as the one given by the robust sandwich estimator in Equation (15) except for the factor $\frac{N}{N-p}$ in Equation (16). We choose to use the above factor in the general expression for the robust sandwich estimator because it reduces to the $HC_1$ definition in the case of linear regression (see SAS PROC REG documentation (SAS Institute Inc. 2018c)). The $HC_1$ definition is preferable in small samples.

We also compared various goodness-of-fit measures and statistical tests discussed in the previous sections with the pooled data analyses. As expected, all measures and tests agreed within machine precision, except for the ROC curves and the Hosmer-Lemeshow statistics. The ROC curve and Hosmer-Lemeshow statistics are dependent on the values specified for

*min_count_per_grp* at each data partner. When *min_count_per_grp is set to 1* at all the data partners, the agreement is within machine precision to standard SAS procedures. Otherwise, we can only approximate these statistics and their agreement is dependent on *min_count_per_grp* $/n_k$.

|  | Distributed Regression | | | Pooled Individual-level Regression | | | Difference in Estimates | Difference in Standard Errors | Difference in Robust Standard Errors |
|---|---|---|---|---|---|---|---|---|---|
| Variable | Parameter Estimate | Standard Error | Robust Standard Error | Parameter Estimate | Standard Error | Robust Standard Error | | | |
| Intercept | 31.79302 | 1.68240 | 1.55065 | 31.79302 | 1.68240 | 1.55065 | -1.09E-12 | 2.40E-14 | -2.49E-14 |
| crim | -0.23283 | 0.04755 | 0.04661 | -0.23283 | 0.04755 | 0.04661 | -1.42E-15 | 1.03E-15 | 3.33E-16 |
| indus | -0.51302 | 0.08165 | 0.07754 | -0.51302 | 0.08165 | 0.07754 | 4.62E-14 | 2.22E-15 | 7.77E-16 |
| dis | -1.05423 | 0.22632 | 0.21689 | -1.05423 | 0.22632 | 0.21689 | 1.17E-13 | 2.72E-15 | -2.66E-15 |
| dummy_dp_var2 | 4.62054 | 0.88611 | 0.76374 | 4.62054 | 0.88611 | 0.76374 | 2.24E-13 | 1.85E-14 | -2.89E-15 |
| dummy_dp_var3 | -1.22053 | 1.04369 | 1.09139 | -1.22053 | 1.04369 | 1.09139 | 9.33E-14 | 3.04E-14 | 1.40E-14 |

**Table 8: Distributed linear regression vs. pooled individual-level linear regression from Example 1 described in Section 3.6.**

|  | Distributed Regression | | | Pooled Individual-level Regression | | | Difference in Estimates | Difference in Standard Errors | Difference in Robust Standard Errors |
|---|---|---|---|---|---|---|---|---|---|
| Parameter | Parameter Estimate | Standard Error | Robust Standard Error | Parameter Estimate | Standard Error | Robust Standard Error | | | |
| Intercept | 1.68778 | 0.53174 | 0.49189 | 1.68778 | 0.53174 | 0.49189 | 6.66E-15 | 2.78E-15 | -1.87E-11 |
| crim | -0.15315 | 0.04653 | 0.04258 | -0.15315 | 0.04653 | 0.04258 | -2.78E-16 | -5.55E-17 | -1.23E-10 |
| indus | -0.10329 | 0.02570 | 0.02383 | -0.10329 | 0.02570 | 0.02383 | -2.78E-16 | 1.21E-16 | -1.99E-13 |
| dis | -0.16344 | 0.07341 | 0.07045 | -0.16344 | 0.07341 | 0.07045 | -9.44E-16 | 3.75E-16 | -4.43E-12 |
| dummy_dp_var2 | 1.33919 | 0.27156 | 0.26679 | 1.33919 | 0.27156 | 0.26679 | 4.44E-16 | 5.55E-17 | -8.69E-13 |
| dummy_dp_var3 | 0.31595 | 0.37325 | 0.38528 | 0.31595 | 0.37325 | 0.38528 | 2.66E-15 | 5.55E-17 | -2.20E-10 |

**Table 9: Distributed logistic regression vs. pooled individual-level logistic regression from Example 2 described in Section 3.6.**

## 5. Discussion

We have developed, tested, and validated a DRA application using SAS software for linear and logistic regression. The application requires only Base SAS and SAS/STAT modules and can be used on any operating system on which SAS can be installed (Windows, Unix, Linux, etc). The DRA application produces results identical, within machine precision, to the results obtained from the corresponding pooled individual-level data analysis with standard SAS procedures (e.g., PROC REG, PROC LOGISTIC). We introduce an approach for residual analysis and approximations for computing an ROC curve, AUC, and Hosmer-Lemeshow statistic that flexibly accommodate data partners' privacy constraints. Our DRA application was successfully tested on several datasets using various optional parameters.

While the theoretical foundation for DRA is well-documented, the implementation of DRA in practice is challenging. The iterative nature DRA requires iterative data exchanges between the analysis center and data partners. These iterations are labor- and resource-intensive and require extensive coordination. We integrated our DRA application into PopMedNet, an open-source distributed networking software that allows automatable iterative file transfer between the analysis center and data partners. An important advantage of using PopMedNet is that all file transfers between data partners and the analysis center are achieved through secure HTTPS/SSL/TLS connections. There are no open ports, Virtual Private Networks, or any external access to data partner or analysis center data. Both PopMedNet DataMart Client and SAS instances are run under the user accounts that the data partners create and maintain. Details regarding integration between our DRA application and PopMedNet can be found elsewhere (Her et al. 2018). While our current implementation of DRA uses PopMedNet, our DRA application can be implemented manually or integrated with any data transferring software that meets the specifications described in Section 3.4.

As described in our companion paper, we have also successfully developed and tested Cox proportional hazards DRA. We plan to extend our DRA implementation to other GLMs used in medical and epidemiological research. In particular, we are interested in extending the application Poisson and negative binomial models for count data and the Gamma model for continuous non-negative outcomes (e.g., length of stay and cost of hospitalization). Future work will also expand the DRA application to vertically partitioned data environments, where different databases contain different information for the same individuals (Reiter et al. 2004; Fienberg et al. 2006; Li et al. 2016).

# Appendix

## A. Main macros used in the implementation of iteratively reweighted least squares algorithm

In this appendix we describe the key macros used by the iteratively reweighted least squares (IRLS) algorithm. The full code of all macros used in our DRA application's packages is available online at https://www.sentinelinitiative.org/sentinel/methods/utilizing-data-various-data-partners-distributed-manner.

*%calc_genmod_vars*. This macro is executed at the data partner and calculates various generalized linear model (GLM) variables at the individual record level. These include the linear predictor $\hat{\boldsymbol{\beta}}^T \mathbf{Z}_{i,k}$, mean $\mu(\boldsymbol{\beta}^T \mathbf{Z}_{i,k})$ and variance $v(\boldsymbol{\beta}^T \mathbf{Z}_{i,k})$, redefined outcome $\tilde{Y}_{i,k}$ (see Equation (11)) and effective weight $\tilde{w}_{i,k}(\boldsymbol{\beta}_m^T)$. The parameter *regr_type_cd* (1: linear, 2: logistic) defines the appropriate expressions for $\mu(\boldsymbol{\beta}^T \mathbf{Z}_{i,k})$, and $v(\boldsymbol{\beta}^T \mathbf{Z}_{i,k})$ in terms of the linear predictor. If one wants to extend the current functionality to other types of GLMs (beyond linear and logistic) one will need to define appropriate expressions for $\mu(\boldsymbol{\beta}^T \mathbf{Z}_{i,k})$, and $v(\boldsymbol{\beta}^T \mathbf{Z}_{i,k})$ in this macro and assign a new *regr_type_cd*.

*%calculate_sscp*. This macro is executed at the data partner. It takes the individual-level dataset as an input with the macro *%calc_genmod_vars* and outputs the dataset with the sums of squares and cross products (SSCP) matrix for a given data partner: $\boldsymbol{SSCP}(\mathbf{Z}_k || \tilde{\mathbf{Y}}_{km}(\boldsymbol{\beta}_m), \tilde{\mathbf{W}}_{km}(\boldsymbol{\beta}_m))$ (see Section 2.2.1). The actual calculations are performed by PROC CORR with the SSCP option.

*%lin_reg_with_sscp_input*. This macro is executed at the analysis center. Before calling this macro, the contributions to the SSCP matrix from all data partners are tranfer to the analysis center and summarized into a single SSCP matrix (see Equation (12)). The resulting SSCP dataset is used as an input into the macro *%lin_reg_with_sscp_input*.. The most important calculations are performed by PROC REG, as explained in Section 2.2.2, which has the ability to accept the SSCP matrix as an input dataset instead of the individual-level dataset. The procedure automatically calculates many desired statistics. These include not only regression estimates $\hat{\boldsymbol{\beta}}$, but also the inverse matrix $(\mathbf{Z}^T \tilde{\mathbf{W}}(\hat{\boldsymbol{\beta}}) \mathbf{Z})^{-1}$ which is used to calculate the covariance matrix $\widehat{cov}(\hat{\boldsymbol{\beta}})$ and standard errors. Note, PROC REG does not have a special option that indicates that an input dataset is a SSCP dataset rather than individual-level dataset. Instead, PROC REG checks if the dataset's metadata has a property TYPE=SSCP. This property has to be explicitly set to SSCP when the dataset is created and before it is used by PROC REG.

*%check_track_convergence*. This macro checks for convergence of the GLM using the relative convergence criteria. When convergence is achieved the macro sets parameter *reg_conv_in*=1, which instructs the programs to switch to the calculation of final statistics.

## B. Goodness-of-fit measures

**Linear regression:**

Let's define:

$\bar{Y} = \frac{1}{N}\sum_{k=1}^{K}\sum_{i=1}^{n_k} Y_{i,k}$: the sample average of the outcome,

$SSE = \sum_{k=1}^{K}\sum_{i=1}^{n_k}(Y_{i,k} - \hat{Y}_{i,k})^2$: the error sum of squares

$SSE_1 = \sum_{k=1}^{K}\sum_{i=1}^{n_k}(Y_{i,k} - \bar{Y})^2$: the total sum of squares corrected for the mean for the dependent variable

The $R^2$ is defined as:

$$R^2 = 1 - \frac{SSE}{SSE_1}$$

Akaike information criterion:

$$AIC = N\ln\left(\frac{SSE}{N}\right) + 2p \qquad (19)$$

with $p$ as the number of model parameters.

Bayesian information criterion is defined as:

$$BIC = N\ln\left(\frac{SSE}{N}\right) + 2(p+2)q - 2q^2$$

$$\text{where } q = \frac{N\hat{\sigma}^2}{SSE}$$

Schwarz's Bayesian criterion is defined as:

$$SBC = N\ln\left(\frac{SSE}{N}\right) + p\ln(N)$$

**Logistic Regression:**

Log likelihood is defined as:

$$LL = \sum_{i,k} Y_{i,k}\ln(\mu_{i,k}) + (1 - Y_{i,k})\ln(1 - \mu_{i,k}) = \sum_{i,k} Y_{i,k}\beta^T Z_{i,k} - \log(1 + \exp(\beta^T Z_{i,k}))$$

Log likelihood ratio statistic for the global null hypothesis test is given by:

$$D = 2(LL - LL(0))$$

where $LL(0)$ is the log likelihood of the model with an intercept only:

$$LL(0) = N \sum_{i,k} \bar{Y} ln(\bar{Y}) + (1 - \bar{Y}) ln(1 - \bar{Y})$$

The generalized coefficient of determination (generalized R-square) is:

$$G_{RSQ} = 1 - \exp\left[\frac{2(LL(0) - LL)}{N}\right]$$

Deviance (relative to the intercept only model) is:

$$D = -2 \sum_{i,k} \left[ Y_{i,k} ln\left(\frac{Y_{i,k}}{\mu_{i,k}}\right) + (1 - Y_{i,k}) ln\left(\frac{1 - Y_{i,k}}{1 - \mu_{i,k}}\right) \right]$$

Akaike information criterion is:

$$AIC = -2LL + 2p \qquad (20)$$

Note that for the linear regression model, Equation (20) does not coincide with the $AIC$ Equation (19) used by PROC REG. However, for the normal likelihood these expressions differ only by a constant, which is irrelevant for model comparison. To be comparable with SAS, we use Equation (19) for linear regression and Equation (20) for logistic regression.

Akaike criterion corrected for finite sample size is defined as:

$$AICC = -2LL + 2p \frac{N}{N - p - 1}$$

Bayesian information criterion is defined as:

$$BIC = -2LL + p \ln(N)$$

## C. Parameters for the main macro *%distributed_regression*.

The table below describes all parameters for the macro *%distributed_regression* used in distributed linear and logistic regression. The parameters for Cox regression will be described elsewhere.

| Parameter | Description |
| --- | --- |
| RunID | Identifier for a given macro call. It is used to form a prefix &RunID for the names of all output datasets. This allows multiple calls of the main DRA macros within the same distributed regression request. Preferably less than 4 characters. *Required.*<br>*Example: RunID=dr1* |
| reg_ds_in | The name of the input analytic dataset. The analytic dataset at each data partner must have the same name and located in the SAS library called DATA_IN (defined in the data partner's SAS wrapper). *Required.*<br>*Example: reg_ds_in=LINEAR_KARR_2005* |
| dp_cd_list | The list of participating data partners separated by space. *Required.*<br>*Example: dp_cd_list=7 15 19* |
| regr_type_cd | Defines the type of regression.<br>1=linear; 2=logistic; 10=Cox. *Required.*<br>*Example: regr_type_cd=1* |
| dependent_vars | Name of the dependent variable in the regression. *Required.*<br>*Example: dependent_vars= medv* |
| independent_vars | List of the independent variables in the regression. *Required.*<br>*Example: independent_vars=crim indus dis* |
| NOINT | When set to NOINT the regression analysis fits without an intercept. Default is blank, which fits the model with an intercept. Not relevant for Cox regression. *Optional.* |

| Freq | Name of the variable that indicates frequency of an observation. *Optional.* *Example: freq=freq_variable* |
|---|---|
| Weight | Name of the variable that indicates weight of an observation. *Optional.* *Example, weight=weight_variable* |
| tbl_intial_est | Name of the table with initial guesses of the regression parameters. This table must have a column for each regression parameter estimates. The column names for the parameter estimates should be the same as the names of the corresponding independent variables, specified in *independent_vars*. (Same structure as special SAS dataset of the TYPE=PARM). If *tbl_initial_est* is not specified, all initial guesses are set to 0. Note, this is different from the default in PROC LOGISTIC, which all initial guesses are 0 except for the intercept, which is the average of the outcome variable. The dataset *tbl_intial_est* should be located in the SAS library *infolder* defined in the wrapper at the analysis center. Not relevant for linear regression. *Optional.* *Example: tbl_intial_est=Model_Coeff_0* |
| xconv | Relative convergence criteria. The same definition as the one used by SAS. *Optional.* Default is 1E-4. |
| max_iter_nb | Maximum number of allowed iterations. *Optional*. Default is 20. |
| alpha | Level of statistical significance. *Optional.* Default is 0.05. |
| groups | Number of groups used in the calculation of residuals summary statistics. *Optional*. Default is 10. |
| wait_time_min | Minimum time interval for checking for the trigger file *files_done.ok*. Measured in seconds. Used by the macro *%file_watcher*. *Optional*. Default is 3. |
| wait_time_max | Maximum time interval for checking for the trigger file *files_done.ok*. Measured in seconds. Used by the macro *%file_watcher*. *Optional*. Default is 7,200, which is 2 hours |

| last_runid_in | If one wants to run more than one regression (can be different regression models) within the same request one should specify *last_runid_in=0* for the first few calls of this macro and to 1 for the last call. *Optional*. Default is 1. |
|---|---|
| test_env_cd | Set to 1 to execute the DRA application in the special development/testing environment. The directory structure in this environment is the same as the structure at the analysis center. When set to 1 the program can be executed within a single SAS session with the code for different data partners running sequentially. It allows testing for most of the SAS code without the need of a data transferring software. *Optional*. Default is 0 which means production environment. |
| max_numb_of_grp | Sets upper limit to the number of groups for summarized data returned to the analysis center. Normally the number of groups is determined by parameters *min_count_per_grp* or *min_count_per_grp_glob*. However, for large datasets this can result in large file sizes transferred from data partners to the analysis center. This is often unnecessary and this parameter puts a cap on the number of rows returned to the analysis center. *Optional*. Default is 10,000. |
| min_count_per_grp_glob | Sets minimum count per cell for summarized data returned to the analysis center. It is only used if a data partner site does not specify parameter *min_count_per_grp* in their master program. This affects datasets used for residual analysis and goodness-of-fit measures (ROC and Hosmer-Lemeshow statistic for logistic regression). *Optional*. Default is 6. |

## D. Final output datasets

Below is a table with a list and description of final output datasets created by the macro %*distributed_regression* for GLMs. All datasets are located in the subdirectory *msoc* at the analysis center. The datasets from a given run have the same prefix equal *&RunID*. For example, for *&RunID=dr1* the *&prefix=dr1*.

| Dataset Name | Dataset Description |
| --- | --- |
| &PREFIX.ANOVA | Has the same structure as the ODS table *ANOVA* generated by PROC REG. Includes standard statistics for analysis of variance: various sum of squares, F-value and corresponding p-value. Used for analysis of linear regression only. |
| &PREFIX.COLLINDIAG | Has the same structure as the ODS table *CollinDiag* generated by PROC REG. Provides collinearity diagnostic between independent variables for linear and non-linear models. |
| &PREFIX.CONVRG_STATUS | Has similar structure as the ODS table *ConvergenceStatus* generated by PROC GENMOD. Also contains information about number of iteration and convergence criteria. Not applicable to linear model. |
| &PREFIX.COV_EST | Has the same structure as the ODS table *CovB* generated by PROC REG and PROC LOGISTIC. Includes information about model-based covariance of estimates. Applicable to all models. |
| &PREFIX.HC_COV | The same structure as the table *&PREFIX.COV_EST* but has the information about covariance of estimates based on the robust sandwich estimator (heteroscedastic covariance). Applicable to all models. |

| Dataset Name | Dataset Description |
|---|---|
| &PREFIX.GLOB_NULL_CHISQ | Has the same structure as the ODS table *GlobalTests* generated by PROC LOGISTIC. Includes Chi-Square statistic, degrees of freedom and p-value for the global null hypothesis test. Not used for linear model which uses F-test (table *&PREFIX.ANOVA*). |
| &PREFIX.HL_CHISQ | Has structure similar to the ODS table *LackFitChiSq* generated by PROC LOGISTIC. Includes information about Hosmer-Lemeshow chi-square test results. |
| &PREFIX.HL_PARTITION | Has the same structure as the ODS table *LackFitPartition* generated by PROC LOGISTIC. Includes information about partition for the Hosmer-Lemeshow test. |
| &PREFIX.INVXPX | Has inverse of the negative Hessian matrix. For linear regression it is the same as inverse of the matrix **X'X**. Applicable to all models. |
| &PREFIX.ITER_PARMS_HIST | Has the same structure as the ODS table *IterHistory* generated by PROC LOGISTIC. Includes information about iteration history. Not applicable to linear model. |
| &PREFIX.MODELFIT | Has structure similar to the ODS table *ModelFit* generated by PROC GENMOD. It has information about various goodness-of-fit measures including AIC, AICC, BIC, R-square. For non-linear regression the R-square represents generalized R-square (coefficient of determination). Applicable to all models. |

| Dataset Name | Dataset Description |
|---|---|
| &PREFIX.MODEL_COEFF | Has the same structure as the output dataset specified by option *OUTEST* in PROC REG/PROC LOGISTIC. Includes information about regression coefficients in the longitudinal form: a single row with a column for each coefficient. |
| &PREFIX.P_EST_HC | Has the same structure as the ODS table *ParameterEstimates* generated by PROC REG. Includes information about regression coefficients in the vertical form with separate row for each coefficient. In addition, it has columns for model and robust standard errors, p-values, upper and lower confidence limits. Applicable to all models. |
| &PREFIX.P_EST | Has the same structure as *&PREFIX.P_EST_HC* but without columns based on robust sandwich estimator. Applicable to all models. |
| &PREFIX.RESID_SUM | Has overall sum and/or mean values for quantities observed and predicted outcome, residuals, square of residuals, log likelihood and various goodness-of-fit measures. Applicable to all models. |

| Dataset Name | Dataset Description |
|---|---|
| &PREFIX.RESID_SUM_BY_PCT | Has summary statistics based on final output dataset at each data partner. The data are grouped by percentiles of the predicted values. The number of observation per group for a data partner can vary slightly due to ties. The number of groups is determined by the macro parameter *groups* specified in the main macro **%*distributed_regression***. The default value is *groups=10*. The summary statistics include mean value for observed and predicted outcome, residuals, square of residuals and model variance. It also includes number of observations per group and a number of distinct values of predicted outcome. The dataset can be used to generate various plots and visually evaluate the goodness-of-fit by the regression model. |
| &PREFIX.RESID_SUM_BY_PCT2 | Has the same structure as *&PREFIX.RESID_SUM_BY_PCT* but with the number of groups determined by the parameter *min_count_per_grp* specified by the data partner. If the parameter *min_count_per_grp* is not specified by a data partner then the parameter *min_count_per_grp_glob*, specified in the macro **%*distributed_regression*** is used. The dataset provides the most detailed level of summarization allowed by each data partner. It is used to approximate the calculation of the ROC curve, AUC, and Hosmer-Lemeshow statistic. It can also be used for residual analysis. Applicable to all models. |
| &PREFIX.ROC | It has the same structure as the output dataset specified by option *OUTROC* in PROC LOGISTIC plus a column with the value of AUC characteristic. Includes data necessary for plotting ROC curve. Applicable to logistic regression only. |

## E. Approximation of the ROC curve, AUC, and Hosmer-Lemeshow statistic

In this appendix, we describe an approximate approach, which uses a pre-summarized dataset from each data partner, to calculate the receiver operating characteristic (ROC) curve, area under the ROC curve (AUC), and Hosmer-Lemeshow statistic. It allows each data partner to control the level of summarization via the parameter *min_count_per_grp*. When this parameter is set to *min_count_per_grp=1* at all data partners, our approach recovers the usual ROC curve, AUC, and Hosmer-Lemeshow statistic computed from individual-level data. In the Section 4.2.2, we show an example of ROC, AUC, and Hosmer-Lemeshow test results for *min_count_per_grp=6* and compare them with the results obtained by the PROC LOGISTIC on combined individual-level data.

The first step is to create a summarized dataset at each data partner, similar to the one we used for residual analysis but using the finest level of summarization allowed by each data partner (see description of the dataset *&prefix0.resid_sum_by_pct2* in Appendix D). Specifically, we choose the number of groups/bins $n_{k,grp}$ for data partner k as:

$$n_{k,grp} = \text{int}\left(\frac{n_k}{\text{min\_count\_per\_grp}}\right) \quad (21)$$

where int() is the integer function.

The summarized datasets from all data partners are brought to the analysis center, combined into a single dataset, and sorted by mean predicted probability. The Table 10 below shows a few records and variables from this dataset.

| PROB | Nobs | Dist_PROB_Cnt_per_bin | RESP__Mean | RESP | NO_RESP |
|---:|---:|---:|---:|---:|---:|
| 0.15502 | 6 | 6 | 0.00000 | 0 | 6 |
| 0.16849 | 6 | 6 | 0.00000 | 0 | 6 |
| 0.17823 | 6 | 6 | 0.16667 | 1 | 5 |
| 0.18998 | 6 | 6 | 0.16667 | 1 | 5 |
| 0.20931 | 6 | 6 | 0.33333 | 2 | 4 |
| 0.21094 | 6 | 5 | 0.33333 | 2 | 4 |

Table 10: Example records from the dataset used to calculate the ROC Curves and Hosmer-Lemeshow statistics

Here, the variable *PROB* represent the average predicted probability for a bin $PROB = \bar{\mu}$, *Nobs* is the total number of observation in a bin, *Dist_PROB_Cnt_per_bin* is the number of distinct predicted probabilities (i.e. $\mu$) per bin, *RESP__Mean* is the mean of the observed response per bin ($RESP\_Mean = \bar{Y}$), *RESP* is the total number of observations with outcome $Y = 1$ in a bin, and *NO_RESP* is the number of observation with outcome $Y = 0$ *(i.e. Nobs–RESP)*.

**Calculation of ROC Curve and AUC statistic**

For ROC calculations, we only need the variables: *PROB*, *RESP*, *NO_RESP*. We apply to the summarized dataset shown in Table 10 the standard ROC algorithm used for individual-level data. The description of this algorithm below follows the PROC LOGISTIC documentation for ROC curve with some modifications to the notations (SAS Institute Inc. 2018b). Suppose we have a dataset with variables *PROB*, *RESP*, and *NO_RESP* where *PROB* is the predicted probability of the outcome with $Y = 1$, *RESP* is the total number of observations with outcome $Y = 1$ in a bin, and *NO_RESP* is the number of observation with outcome $Y = 0$ (Table 10). Let's $l$ be a row index in the dataset then:

Number of correctly predicted event responses:

$$POS(z) = \sum_l RESP_l * I(PROB_l \geq z)$$

Number of correctly predicted nonevent responses:

$$NEG(z) = \sum_l NO\_RESP_l * I(PROB_l < z)$$

Number of falsely predicted event responses:

$$FALPOS(z) = \sum_l NO\_RESP_l * I(PROB_l \geq z)$$

Number of falsely predicted nonevent responses

$$FALNEG(z) = \sum_l RESP_l * I(PROB_l < z)$$

Sensitivity of the test:

$$SENSIT(z) = \frac{POS(z)}{Total\_POS}$$

One minus the specificity of the test:

$$1MSPEC(z) = \frac{FALPOS(z)}{Total\_NEG}$$

The ROC curve is a plot of sensitivity (*SENSIT*) against 1–specificity (*1MSPEC*). The AUC is the area under the ROC curve. After sorting the data by descending $PROB_{l-1}$ (increasing $1MSPEC$) it can be calculated using trapezoidal integration rule:

$$AUC = 0.5 \sum_{l} [SENSIT(\text{PROB}_l) + SENSIT(\text{PROB}_{l-1})] * [1MSPEC(\text{PROB}_l) - 1MSPEC(\text{PROB}_{l-1})]$$

**The calculation of the Hosmer-Lemeshow statistic**

We can calculate the Hosmer-Lemeshow statistic using the same summarized dataset (Table 10) that we used to approximate the ROC and AUC calculation. However, it the Hosmer-Lemeshow statistic is more sensitive to the original ordering of data by *PROB* than the AUC. We are able to get a better approximation for the Hosmer-Lemeshow statistic when we use the variable *Dist_PROB_Cnt_per_bin* to transform the above summarized dataset into a new dataset, which more closely resembles the ordering of *PROB* in the individual-level datasets left at each data partner site. Specifically, for each row with *Nobs* and *Dist_PROB_Cnt_per_bin* we create the following set of records:

a) one record with the same value of *PROB* as the original record and $Nobsnew = Nobs - Dist\_PROB\_Cnt\_per\_bin + 1$

b) $Dist\_PROB\_Cnt\_per\_bin - 1$ records with consecutive values of *PROB* differing from one another by a very small amount, e.g., $10^{-10}$, are created. Below is an example of such a transformation for a record with $Nobs = 6$, $N\_Dist\_Prob = 4$, and $PROB = 0.2109439184$. This single record is transformed into the following 5 records:

| PROB | Nobsnew | RESP_Mean |
|---|---|---|
| 0.2109439184 | 2 | 0.33333 |
| 0.2109439185 | 1 | 0.33333 |
| 0.2109439186 | 1 | 0.33333 |
| 0.2109439187 | 1 | 0.33333 |
| 0.2109439188 | 1 | 0.33333 |

In the above example the values of *PROB* differ only in the last 10[th] digit. Adding such a small amount to *PROB* has virtually no effect on arithmetic operations but it does affect ordering of records across data partners and the way groups are formed by Hosmer-Lemeshow algorithm.

We now apply the standard Hosmer-Lemeshow algorithm to this transformed dataset. This involves creating $g$ groups of approximately equal size (see algorithm in Appendix F) and calculating the averages $\text{MEAN(PROB)} = \bar{\mu}_l$ and $\text{MEAN(RESP\_Mean)} = \bar{Y}_l$ within each group $l$. The Hosmer-Lemeshow statistic is calculated as:

$$\chi^2_{HL} = \sum_{l=1}^{g} \frac{(\bar{Y}_l - \bar{\mu}_l)^2}{\bar{\mu}_l(1 - \bar{\mu}_l)}$$

The number of groups $g$ can be changed using the macro parameter *groups* in the main macro **%*distributed_regression***. The default value is 10 which is what SAS uses in PROC LOGISTIC. The p-value of the Hosmer-Lemeshow statistic is calculated using a Chi-square distribution with $g - 2$ degrees of freedom. Large values of $\chi^2_{HL}$ (and small p-values) indicate a lack of fit of the model (see Hosmer-Lemeshow documentation in SAS documentation for PROC LOGISTIC) (SAS Institute Inc. 2018a).

**Table 7** of the main manuscript provides results obtained using this procedure with *min_count_per_grp=6*. The parameter *groups* is set to the default value of 10. For comparison, we also provided the result obtained using PROC LOGISTIC on combined individual-level dataset in the same table. The results are fairly close. It is interesting to note that without the above procedure to create multiple rows using variable *Dist_PROB_Cnt_per_bin* the Chi-Square for the Hosmer-Lemeshow statistic and p-value would be 18.1 and 0.02, respectively. This differs significantly more from the results for PROC LOGISTIC.

### F. Algorithm for grouping data into bins based on percentiles of predicted outcome

The approach that we use to create groups by the predicted outcome $\mu$ is similar to the one used by SAS in the computation of the Hosmer-Lemeshow statistic (SAS Institute Inc. 2018a). We extended it to any number of groups and added a constraint to the minimum allowed number of observations per group, *min_count_per_grp*. The variables $\mu_j$, $f_j$ introduced below correspond to the columns *PROB* and *RESP* in the dataset shown in Table 10, the variable $n_{min}$ corresponds to the macro parameter *min_count_per_grp*, the variable $n_{grp}$ corresponds to the parameter *groups* when calculating summary dataset *prefix0.resid_sum_by_pct* and to the expression for $n_{grp}$ in the Equation (21) when calculating the dataset *prefix0.resid_sum_by_pct2*.

Let $N_{total}$ be the total number of observations in a dataset with individual observations of predicted outcome $\mu_j$, $n_{grp}$ is the number of desired groups, $n_{min}$, is the minimum allowed number of observations per group. We define the target number of observations for each group $M_{target}$ as:

$$M_{target} = \max\left(\text{int}\left(\frac{N_{total}}{n_{grp}} + 0.5\right), n_{min}\right)$$

Here int() is the integer function.

Algorithm:

1) Create a summarized dataset with one record per distinct value of $\mu_j$ and the count of observations $f_j$ per $\mu_j$. Sort the dataset by $\mu_j$ in increasing order. The purpose of this step is to ensure that observations with the same value of $\mu_j$ are not split into separate groups.

2) Start reading records one by one and calculate cumulative count of observations for a given group $f_{grp\_cum,j}$.

3) Assign the first observation of $\mu_j$ to the first group and continue to assign subsequent values to that group as long as:

$$f_{grp\_cum,j} < M_{target} \text{ and } f_{grp\_cum,j} + f_j/2 \leq M_{target}$$

4) Otherwise put $\mu_j$ into the next group and reset $f_{cum,j} = f_j$. Repeat steps 2 and 3 to assign $\mu_j$ into subsequent groups.
5) If $f_{cum,j}$ in the last group does not exceed $M_{target}/2$ then the last two groups are collapsed to form only one group.

Note that the actual number of groups created can be smaller than the number of requested groups $n_{grp}$ if the number of distinct values of $\mu_j$ is smaller than $n_{grp}$.